\documentclass[11pt]{article}
\usepackage{graphicx}
\usepackage{amsmath}
\usepackage{amssymb}
\usepackage{amsxtra}
\usepackage{bm}

\title{Structure of quantum corrections in ${\cal N}=1$ supersymmetric gauge theories}
\author{K.V. Stepanyantz\\Moscow State University, Physical Faculty, Department of \\Theoretical Physics, 119991, Moscow, Russia,\\ stepan@m9com.ru}

\begin{document}
\maketitle

\begin{abstract}
Some recent research of quantum corrections in ${\cal N}=1$ supersymmetric theories is briefly reviewed. The most attention is paid to the theories regularized by higher covariant derivatives. In particular, we discuss, how the NSVZ and NSVZ-like relations appear with this regularization and how one can construct the NSVZ scheme in all orders.
\end{abstract}


\section{Introduction}\label{s:intro}
\hspace{\parindent}

${\cal N}=1$ sypersymmetric extensions of the Standard Model (SM) are very interesting candidates for describing physics beyond it \cite{Mohapatra:1986uf}. In these theories there are no quadratically divergent quantum corrections to the Higgs mass, the running of coupling constants agrees with the predictions of the Grand Unified Theories, and the proton lifetime (proportional to $M_X^4$) is much larger than in the non-supersymmetric case. This makes them very attractive from the phenomenological point of view. However, the supersymmetric extensions of SM predict a lot of new particles, which are superpartners of quarks, leptons, gauge bosons and Higgs bosons. Supersymmetry also requires two Higgs doublets, which produces $2\times 2\times 2 - 3 = 5$ Higgs bosons. To make masses of superpartners sufficiently large, it is necessary to break supersymmetry. Although it is highly desirable to break supersymmetry spontaneously, the simplest models (like MSSM) include soft terms, which explicitly break supersymmetry, but do not produce quadratic divergences. Investigation of quantum corrections in supersymmetric theories and theories with softly broken sypersymmetry and comparing them with experimental data can provide information about physics beyond SM.

It is convenient to describe ${\cal N}=1$ supersymmetric theories in ${\cal N}=1$ superspace, because in this case supersymmetry is a manifest symmetry. In this language, the renormalizable ${\cal N}=1$ SYM theory (with a simple gauge group $G$, for simplicity) is described by the action

\begin{eqnarray}\label{SYM_Superfield_Action}
&& S = \frac{1}{2 e_0^2}\,\mbox{Re}\,\mbox{tr}\int d^4x\,
d^2\theta\,W^a W_a + \frac{1}{4} \int d^4x\, d^4\theta\,\phi^{*i}
(e^{2V})_i{}^j \phi_j\nonumber\\
&& + \Big\{\int d^4x\,d^2\theta\,\Big(\frac{1}{4} m_0^{ij} \phi_i
\phi_j + \frac{1}{6}\lambda_0^{ijk} \phi_i \phi_j \phi_k\Big) +
\mbox{c.c.}\Big\},\nonumber
\end{eqnarray}

\noindent
where $\theta$ denotes auxiliary Grassmannian coordinates. The real superfield $V(x,\theta,\bar\theta)$ is the gauge superfield, and the supersymmetric gauge field strength is defined as $W_a = \bar D^2 \left(e^{-2V} D_a e^{2V}\right)/8$. The matter superfields $\phi_i$ are chiral, $\bar D_{\dot a}\phi_i = 0$, where
in our notation $D_a$ and $\bar D_{\dot a}$ denote the right and left supersymmetric covariant derivatives, respectively. In terms of superfields the gauge transformations can be written as

\begin{equation}
\phi \to e^{A}\phi;\qquad e^{2V} \to e^{-A^+} e^{2V} e^{-A},
\end{equation}

\noindent
and are parameterized by a chiral superfield $A = i e_0 A^B T^B$.

Quantum behaviour of sypersymmetric theories is better than in the non-supersymmetric case. For example, in the most interesting for phenomenology case of ${\cal N}=1$ supersymmetry, there are no divergent quantum corrections to the superpotential \cite{Grisaru:1979wc}. Consequently, the renormalization of masses and Yukawa couplings in such theories is related to the renormalization of the chiral matter superfields. As a non-renormalization theorem one can also consider a relation between the $\beta$-function and the anomalous dimensions of the chiral matter superfields which takes place in ${\cal N}=1$ supersymmetric theories \cite{Novikov:1983uc,Jones:1983ip,Novikov:1985rd,Shifman:1986zi},

\begin{equation}\label{NSVZ_Beta_Function}
\beta(\alpha,\lambda) = - \frac{\alpha^2\Big(3 C_2 - T(R) + C(R)_i{}^j \gamma_j{}^i(\alpha,\lambda)/r\Big)}{2\pi(1-C_2\alpha/2\pi)}.
\end{equation}

\noindent
In our notation $r = \mbox{dim}\, G$, and $T^A$ are the generators of the representation $R$ to which the chiral matter superfields belong, such that
$\mbox{tr}\,(T^A T^B) = T(R)\,\delta^{AB}$ and $(T^A T^A)_i{}^j \equiv C(R)_i{}^j$. For the adjoint representation $T(Adj) =C_2$, where $f^{ACD} f^{BCD} \equiv C_2
\delta^{AB}$. The relation (\ref{NSVZ_Beta_Function}) is called the exact NSVZ $\beta$-function, because for the pure ${\cal N}=1$ SYM theory it gives the exact expression for the $\beta$-function. In this paper (following Ref. \cite{Stepanyantz:2016gtk}) we will also discuss the relation between the NSVZ $\beta$-function and the non-renormalization theorem for the triple gauge-ghost vertices. This theorem claims that in ${\cal N}=1$ SYM theories three-point vertices with two ghost legs and one leg of the quantum gauge superfield are finite.

Although a lot of general arguments can be used for obtaining Eq. (\ref{NSVZ_Beta_Function}), see, e.g., \cite{Shifman:1999mv,ArkaniHamed:1997mj,Kraus:2002nu}, it is not so trivial to establish how the NSVZ relation appears in perturbative calculations. Certainly, for doing such calculations the theory should be properly regularized, and the way of removing divergences should be specified. By other words, it is necessary to fix a subtraction scheme. The calculations done with the dimensional reduction \cite{Siegel:1979wq} in the $\overline{\mbox{DR}}$-scheme in the three- and four-loop approximations \cite{Avdeev:1981ew,Jack:1996vg,Jack:1996cn,Harlander:2006xq} demonstrated that Eq. (\ref{NSVZ_Beta_Function}) does not take place starting from the three-loop approximation. However, one can explain the disagreement by the scheme dependence of the NSVZ relation \cite{Kutasov:2004xu,Kataev:2014gxa}. A possibility of this explanation is non-trivial due to some scheme-independent consequences of the NSVZ relation \cite{Kataev:2014gxa,Kataev:2013csa}. Thus, with the dimensional reduction the NSVZ equation is obtained by a special tuning of the subtraction scheme in every order, while the general all-order prescription giving the NSVZ scheme is absent.

Also it should be noted that the dimensional reduction is not mathematically consistent \cite{Siegel:1980qs}, and can break supersymmetry in higher orders \cite{Avdeev:1982np,Avdeev:1982xy}. That is why the use of other regularizations is also reasonable and interesting. In this paper we will mostly discuss various application of the Slavnov higher covariant derivative regularization \cite{Slavnov:1971aw,Slavnov:1972sq} to calculating quantum corrections in ${\cal N}=1$ supersymmetric theories. Unlike the dimensional reduction, this regularization is consistent and can be formulated in a manifestly ${\cal N}=1$ supersymmetric way \cite{Krivoshchekov:1978xg,West:1985jx}. It is also applicable to theories with ${\cal N}=2$ supersymmetry \cite{Krivoshchekov:1985pq,Buchbinder:2014wra,Buchbinder:2015eva}. The main idea of this regularization is to add a term with higher degrees of covariant derivatives to the action of a theory. Then divergences beyond the one-loop approximation disappear, while the remaining one-loop divergences are regularized by inserting the Pauli--Villars determinants into the generating functional \cite{Slavnov:1977zf}. In this paper we will demonstrate that this regularization allows to reveal some interesting features of quantum corrections in supersymmetric theories which are missed in the case of using the dimensional technique.

\section{NSVZ relation in ${\cal N}=1$ SQED}\label{s:sqed}

\subsection{Higher derivative regularization in the Abelian case}
\hspace{\parindent}

We will start with the simplest ${\cal N}=1$ supersymmetric gauge theory, namely, the ${\cal N}=1$ supersymmetric electrodynamics (SQED) with $N_f$ flavors. In the massless case this theory is described by the action

\begin{equation}\label{SQED_Superfield_Action}
S = \frac{1}{4e_0^2}\mbox{Re}\int d^4x\,d^2\theta\,W^a W_a +
\sum\limits_{f=1}^{N_f} \frac{1}{4} \int
d^4x\,d^4\theta\,\Big(\phi_f^* e^{2V}\phi_f + \widetilde\phi_f^*
e^{-2V} \widetilde\phi_f\Big),
\end{equation}

\noindent
which is written in terms of ${\cal N}=1$ superfields. In this formalism supersymmetry is a manifest symmetry of the theory. The usual gauge field is now a component of the real gauge superfield $V$. The terms containing the chiral matter superfields $\phi_f$ and $\widetilde\phi_f$ produce Dirac fermions and the other terms needed for supersymmetry invariance. In the Abelian case the supersymmetric gauge field strength is described by the chiral spinor superfield $W_a = \bar D^2 D_a V/4$. For the theory (\ref{SQED_Superfield_Action}) the NSVZ $\beta$-function (\ref{NSVZ_Beta_Function}) takes the form \cite{Vainshtein:1986ja,Shifman:1985fi}

\begin{equation}
\beta(\alpha) = \frac{\alpha^2 N_f}{\pi}\Big(1-\gamma(\alpha)\Big).
\end{equation}

To regularize the theory (\ref{SQED_Superfield_Action}) by the Slavnov higher derivatives method, we add the term

\begin{equation}
S_\Lambda = \frac{1}{4 e_0^2} \mbox{Re}\,\int d^4x\,d^2\theta\,W^a \Big(R(\partial^2/\Lambda^2)-1\Big) W_a
\end{equation}

\noindent
to the classical action, where the function $R(\partial^2/\Lambda^2)$ contains higher degrees of derivatives. Note that for Abelian theories one should use the usual derivatives (instead of the covariant ones). In the simplest case it is possible to choose $R = 1 + \partial^{2n}/\Lambda^{2n}$. Due to the presence of the higher derivative term, the propagator of the gauge superfield contains higher degrees of the momentum in the denominator, and all diagrams beyond the one-loop approximation become finite. For removing the remaining one-loop divergences, following Ref. \cite{Slavnov:1977zf}, we insert into the generating functional the Pauli--Villars determinants,

\begin{equation}
Z = \int D\mu\,\smash{\prod\limits_I} \Big(\det PV(V,M_I)\Big)^{N_f c_I} \exp\Big\{i S_{\mbox{\scriptsize reg}}
+iS_{\mbox{\scriptsize gf}} + iS_{\mbox{\scriptsize Sources}} \Big\},
\end{equation}

\noindent
with the constants $c_I$ satisfying the conditions $\sum_I c_I = 1$; $\sum_I c_I M_I^2 = 0$. Here $M_I = a_I \Lambda$ (where $a_I$ are constants independent of $\alpha_0$) are masses of the Pauli--Villars superfields proportional to the parameter $\Lambda$ which enters the regulator function $R$.

Below we will see that the NSVZ equation follows from the underlying relation between the two-point Green functions. In ${\cal N}=1$ SQED these two-point Green functions are related to the corresponding part of the effective action by the equation

\begin{eqnarray}
&& \Gamma^{(2)} = \int \frac{d^4p}{(2\pi)^4}\,d^4\theta\,\Big(-
\frac{1}{16\pi} V(-p)\,\partial^2\Pi_{1/2} V(p)\,
d^{-1}(\alpha_0,\Lambda/p)\qquad\nonumber\\
&& + \frac{1}{4} \smash{\sum\limits_{f=1}^{N_f}}
\Big(\phi_f^*(-p,\theta) \phi_f(p,\theta) +
\widetilde\phi_f^*(-p,\theta) \widetilde\phi_f(p,\theta)\Big)
\,G(\alpha_0,\Lambda/p)\Big).
\end{eqnarray}

\noindent
Here $\partial^2\Pi_{1/2}\equiv - D^a \bar D^2 D_a/8$ is a supersymmetric transversal projection operator, and the transversality of the gauge superfield two-point function follows from the Slavnov--Taylor identities.

The function $d^{-1}$ expressed in terms of the renormalized coupling constant $\alpha(\alpha_0,\Lambda/\mu)$ should be finite
in the limit $\Lambda\to \infty$. The charge renormalization constant $Z_3$ is then defined as $Z_3(\alpha,\Lambda/\mu) \equiv \alpha/\alpha_0$.
To construct the renormalization constant $Z$ for the chiral matter superfields, we require finiteness of the function $Z(\alpha, \Lambda/\mu)G(\alpha_0,\Lambda/p)$ in the limit $\Lambda\to\infty$.

According to \cite{Kataev:2013eta}, it is important to distinguish the renormalization group functions (RGF) defined in terms of the bare coupling constant and the ones defined in terms of the renormalized coupling constant. In terms of the bare coupling constant RGF are defined by the equations

\begin{equation}
\beta(\alpha_0) \equiv
\frac{d \alpha_0}{d\ln\Lambda}\Big|_{\alpha=\mbox{\scriptsize
const}};\qquad
\gamma(\alpha_0) \equiv -\frac{d \ln Z}{d\ln\Lambda}\Big|_{\alpha=\mbox{\scriptsize const}}.
\end{equation}

\noindent
They are independent of a renormalization prescription for a fixed regularization, see, e.g., \cite{Kataev:2013eta}, but depend on the regularization. Below we will see that for the theory (\ref{SQED_Superfield_Action}) these RGF satisfy the NSVZ relation in all loops in the case of using the above described version of the higher derivative regularization.

\subsection{Charge renormalization in the lowest loops}
\hspace{\parindent}

Explicit calculations in the lowest loops made with the higher covariant derivative regularization demonstrated that loop integrals giving a $\beta$-function defined in terms of the bare coupling constant are integrals of total derivatives \cite{Soloshenko:2003nc}. They can be also presented as integrals of double total derivatives \cite{Smilga:2004zr}. The $\beta$-function of ${\cal N}=1$ SQED with $N_f$ flavours, regularized by higher derivatives, is calculated by the help of the equation

\begin{eqnarray}
&& \frac{\beta(\alpha_0)}{\alpha_0^2} =
\frac{d}{d\ln \Lambda}\,
\Big(d^{-1}(\alpha_0,\Lambda/p)-\alpha_0^{-1}\Big)\Big|_{p=0}.
\end{eqnarray}

\noindent
By other words, we calculate the two-point Green function of the gauge superfield and differentiate it with respect to $\ln\Lambda$ in the limit of the vanishing external momentum. For example, the two-loop result for the $\beta$-function written as the integral of double total derivatives has the form

\begin{eqnarray}\label{3L_Beta}
&&\hspace*{-8mm} \frac{\beta(\alpha_0)}{\alpha_0^2}
= 2\pi N_f \frac{d}{d\ln\Lambda}
\int \frac{d^4q}{(2\pi)^4}
\frac{\partial}{\partial q^\mu}
\frac{\partial}{\partial q_\mu} \Big\{\sum\limits_{I} c_I \frac{\ln(q^2+M_I^2)}{q^2} + \int\frac{d^4k}{(2\pi)^4}
\nonumber\\
&&\hspace*{-8mm} \times  \frac{2 e^2}{k^2
R_k} \Big(\frac{1}{q^2 (k+q)^2}
- \sum\limits_I c_I \frac{1}{(q^2+M_I^2)((k+q)^2 + M_I^2)}\Big)
\Big\} + O(e^4).
\end{eqnarray}

\noindent
The (essentially larger) three-loop expression can be found, e.g., in \cite{Stepanyantz:2012zz}. Note that the $\beta$-function does not vanish because of integrand singularities. This can be illustrated by a simple example: consider a nonsingular function $f(q^2)$ rapidly decreasing at infinity. Then

\begin{equation}
\int \frac{d^4q}{(2\pi)^4} \frac{\partial}{\partial q^\mu} \Big(\frac{q^\mu}{q^4} f(q^2)\Big) = -\frac{1}{8\pi^2} f(0).
\end{equation}

\noindent
Doing similar calculations it is possible to decrease the number of integrations in Eq. (\ref{3L_Beta}) and reduce this expression to the integral giving the one-loop anomalous dimension of the matter superfield (also defined in terms of the bare coupling constant),

\begin{equation}
\frac{\beta(\alpha_0)}{\alpha_0^2} = \frac{N_f}{\pi}\Big(1-\frac{d}{d\ln\Lambda} \ln
G(\alpha_0,\Lambda/q)\Big|_{q=0}\Big) = \frac{N_f}{\pi}\Big(1-\gamma(\alpha_0)\Big).
\end{equation}

\subsection{NSVZ relation in all loops}
\hspace{\parindent}

The all-loop derivation of the NSVZ relation for RGF defined in terms of the bare coupling constant by the direct summing of supergraphs for ${\cal N}=1$ SQED regularized by higher derivatives has been made in \cite{Stepanyantz:2011jy,Stepanyantz:2014ima} and verified at the three-loop level in \cite{Kazantsev:2014yna}. Here we briefly explain the main ideas of the method of Ref. \cite{Stepanyantz:2011jy}.

First, it is necessary to prove that all loop integrals for the $\beta$-function  defined in terms of the bare coupling constant are integrals of double total derivatives. For this purpose it is convenient to use the background field method which (in the Abelian case) is introduced by making the replacement $V \to V+\bm{V}$, where $\bm{V}$ is the background gauge superfield, in the action. Then we make the formal substitution $\bm{V}\to \theta^4$, after which

\begin{equation}\label{SQED_Bare_Beta}
\frac{d\Delta\Gamma^{(2)}_{\bm{V}}}{d\ln\Lambda}\Big|_{\bm{V}=\theta^4} = \frac{1}{2\pi}{\cal V}_4\cdot \frac{d}{d\ln\Lambda} \Big(d^{-1}(\alpha_0,\Lambda/p)-\alpha_0^{-1}\Big) = \frac{1}{2\pi}{\cal V}_4 \cdot \frac{\beta(\alpha_0)}{\alpha_0^2},
\end{equation}

\noindent
where ${\cal V}_4$ is the (properly regularized) volume of the space-time.

For ${\cal N}=1$ SQED the functional integrals over the matter superfield are Gaussian and can be calculated exactly. This allows operating with some expressions valid in all loops. In particular, it is possible to find the formal expression for the two-point function of the background gauge superfield. Then after the substitution $\bm{V}\to \theta^4$ we try to present the result as an integral of double total derivatives. In the coordinate representation an integral of a total derivative is written as

\begin{equation}
\mbox{Tr} \Big([x^\mu, \mbox{Something}]\Big) -\mbox{Singularities} = -\mbox{Singularities}.
\end{equation}

\noindent
After some non-trivial transformations the result for the expression (\ref{SQED_Bare_Beta}) can be presented as a trace of double commutator, i.e. as an integral of a double total derivative. The details of this calculation are described in Ref. \cite{Stepanyantz:2011jy}. The result does not vanish due to singularities of the integrand, which can be summed in all orders. This gives

\begin{equation}
\frac{d\Delta\Gamma^{(2)}}{d\ln\Lambda}\Big|_{\bm{V}=\theta^4} =
\frac{N_f}{2\pi^2} {\cal V}_4 \Big(1-\frac{d\ln G}{d\ln\Lambda}\Big|_{q=0}\Big) = \frac{N_f}{2\pi^2} {\cal V}_4 \Big(1-\gamma(\alpha_0)\Big),
\end{equation}

\noindent
and we obtain the exact all-order result

\begin{equation}\label{NSVZ_SQED}
\frac{\beta(\alpha_0)}{\alpha_0^2} = \frac{N_f}{\pi} \Big(1-\gamma(\alpha_{0})\Big).
\end{equation}

\noindent
Note that this equation is valid for an arbitrary renormalization prescription in the case of using the higher derivative regularization, because RGF entering it are defined in terms of the bare coupling constant.

In graphical language, this result can be explained as follows \cite{Smilga:2004zr} (see also \cite{Pimenov:2006cu}): If we have a supergraph without external lines, then a contribution to the $\beta$-function can be constructed by attaching two external lines of the background gauge superfield $\bm{V}$ to it, while a contribution to the anomalous dimension is obtained by cutting matter lines in the considered supergraph. The equation (\ref{NSVZ_SQED}) relates both these contributions.

\subsection{How to construct the NSVZ scheme in ${\cal N}=1$ SQED}
\hspace{\parindent}

Eq. (\ref{NSVZ_SQED}) is valid for RGF defined in terms of the bare coupling constant. However, RGF are standardly defined by a different way, in terms of the renormalized
coupling constant,

\begin{equation}
\widetilde\beta(\alpha) \equiv \frac{d\alpha}{d\ln\mu}\Big|_{\alpha_0=\mbox{\scriptsize const}};\qquad
\widetilde\gamma(\alpha) \equiv \frac{d\ln Z}{d\ln\mu}\Big|_{\alpha_0=\mbox{\scriptsize const}},
\end{equation}

\noindent
and are scheme-dependent. However, both definitions of RGF give the same functions, if the conditions

\begin{equation}\label{Boundary_Z}
Z_3(\alpha,x_0) = 1;\qquad Z(\alpha,x_0)=1
\end{equation}

\noindent
are imposed on the renormalization constants, in which $x_0$ is a fixed value of $x = \ln\Lambda/\mu$ \cite{Kataev:2014gxa,Kataev:2013csa,Kataev:2013eta}:
$\widetilde\beta(\alpha_0) = \beta(\alpha_0)$; $\widetilde\gamma(\alpha_0) = \gamma(\alpha_0)$.

$\widetilde\beta$ and $\widetilde\gamma$ are scheme-dependent and satisfy the NSVZ equation only in a certain (NSVZ) scheme. Now, from Eq. (\ref{NSVZ_SQED}) and the above arguments it is evident that for the theory regularized by higher derivatives this NSVZ scheme is fixed in all loops by the boundary conditions (\ref{Boundary_Z}).

The general statements discussed above can be verified by explicit calculations in the lowest loops. They are non-trivial starting from the three-loop approximation, because the $\beta$-function and the anomalous dimension are scheme-dependent starting from the three- and two-loop order, respectively.

For the higher derivative regulator $R_k = 1+k^{2n}/\Lambda^{2n}$

\begin{eqnarray}\label{Alpha_SQED}
&&\hspace*{-4mm} \frac{1}{\alpha_0} = \frac{1}{\alpha} - \frac{N_f}{\pi} \Big(
\ln\frac{\Lambda}{\mu} + b_1\Big) - \frac{\alpha
N_f}{\pi^2} \Big(\ln\frac{\Lambda}{\mu} + b_2\Big) -
\frac{\alpha^2 N_f}{\pi^3}\Big(\frac{N_f}{2}\ln^2
\frac{\Lambda}{\mu}\qquad\nonumber\\
&&\hspace*{-4mm}  - \ln \frac{\Lambda}{\mu}\Big(N_f \sum\limits_{I} c_I \ln
a_I + N_f +\frac{1}{2} -N_f b_1 \Big) +
b_3\Big) + O(\alpha^3);\qquad\\
\label{Z_SQED}
&&\hspace*{-4mm} Z = 1 +
\frac{\alpha}{\pi}\Big(\ln\frac{\Lambda}{\mu}+g_1\Big)
+\frac{\alpha^2 (N_f+1)}{2\pi^2}\ln^2\frac{\Lambda}{\mu}
-\frac{\alpha^2}{\pi^2} \ln\frac{\Lambda}{\mu}
\nonumber\\
&&\hspace*{-4mm} \times  \Big(N_f
\sum\limits_{I} c_I\ln a_I - N_f b_1 + N_f +
\frac{1}{2} - g_1\Big) + \frac{\alpha^2
g_2}{\pi^2} + O(\alpha^3),\qquad
\end{eqnarray}

\noindent
where $b_i$ and $g_i$ are arbitrary finite constants, which fix a subtraction scheme. Differentiating Eqs. (\ref{Alpha_SQED}) and (\ref{Z_SQED}) with respect to $\ln\Lambda$ we construct RGF defined in terms of the bare coupling constant,

\begin{eqnarray}
&&\hspace*{-11mm} \frac{\beta(\alpha_0)}{\alpha_0^2} =
\frac{N_f}{\pi} + \frac{\alpha_0 N_f}{\pi^2} - \frac{\alpha_0^2
N_f}{\pi^3}\Big(N_f \sum\limits_{I} c_I \ln a_I + N_f +
\frac{1}{2} \Big) +
O(\alpha_0^3);\\
&&\hspace*{-11mm} \gamma(\alpha_0) = -\frac{\alpha_0}{\pi} +
\frac{\alpha_0^2}{\pi^2}\Big(N_f \sum\limits_{I} c_I \ln a_I +
N_f + \frac{1}{2}\Big) + O(\alpha_0^3),
\end{eqnarray}

\noindent
which appear to be independent of the constants $b_i$ and $g_i$ and to satisfy the NSVZ relation. However, RGF defined in terms of the renormalized coupling constant,

\begin{eqnarray}\label{BetaR}
&&\hspace*{-12mm}
\frac{\widetilde\beta(\alpha)}{\alpha^2} =
\frac{N_f}{\pi} + \frac{\alpha N_f}{\pi^2} - \frac{\alpha^2 N_f}{\pi^3}\Big( N_f\sum\limits_{I} c_I \ln a_I +N_f
+\frac{1}{2} + N_f (b_2 - b_1) \Big)\nonumber\\
&&\hspace*{-12mm} + O(\alpha^3);\\
\label{GammaR}
&&\hspace*{-12mm} \widetilde\gamma(\alpha) =
-\frac{\alpha}{\pi} + \frac{\alpha^2}{\pi^2}\Big(N_f +
\frac{1}{2}+ N_f \sum\limits_{I} c_I \ln a_I  - N_f
b_1 + N_f g_1\Big) + O(\alpha^3)
\end{eqnarray}

\noindent
depend on these constants and, therefore, on a subtraction scheme. This subtraction scheme can be fixing, e.g., by imposing the conditions (\ref{Boundary_Z}). Choosing $x_0 = 0$, from these equations we obtain $g_2=b_1=b_2=b_3=0$. Therefore, in this scheme only powers of $\ln\Lambda/\mu$ are included into the renormalization constants, while all finite constants vanish. Thus, the considered scheme looks very similar to the minimal subtractions. However, now we use the higher derivative regularization, so that it is reasonable to call this scheme $\mbox{HD}+\mbox{MSL}$, where MSL is the abbreviation for Minimal Subtraction of Logarithms. Substituting the above values of the finite constants into Eqs. (\ref{BetaR}) and (\ref{GammaR}), it is easy to see that in this scheme these RGF satisfy the NSVZ relation.

\subsection{Quantum corrections with the dimensional reduction}
\hspace{\parindent}

It is well known \cite{Avdeev:1981ew,Jack:1996vg,Jack:1996cn} that in the $\overline{\mbox{DR}}$-scheme the NSVZ relation is not valid starting from the three-loop approximation. However, to obtain it, one can specially tune a subtraction scheme in each order. It is also possible to try making calculations similarly to the higher derivative case \cite{Aleshin:2015qqc,Aleshin:2016rrr}. However, the corresponding relation between the functions $d^{-1}$ and $G$ (which is at present obtained only in the lowest orders) has a more complicated form, than for the higher derivative case. The boundary conditions analogous to (\ref{Boundary_Z}) can also be written, but the right hand side of one of them is a series in $\alpha$. It was demonstrated that such a structure agrees with the results obtained in \cite{Jack:1996vg,Jack:1996cn}.

\subsection{NSVZ-like relation in softly broken ${\cal N}=1$ SQED regularized by higher derivatives}
\hspace{\parindent}

NSVZ-like relations \cite{Hisano:1997ua,Jack:1997pa,Avdeev:1997vx} also exist in theories with softly broken supersymmetry for renormalization of the gaugino mass. Their origin is the same as in the case of rigid theories. For example, the exact equation describing the renormalization of the photino mass in softly broken ${\cal N}=1$ SQED,
regularized by higher derivatives,

\begin{equation}
\gamma_m(\alpha_0) = \frac{\alpha_0 N_f}{\pi}\Big[1- \frac{d}{d\alpha_0}\Big(\alpha_0\gamma(\alpha_0)\Big)\Big],
\end{equation}

\noindent
is obtained by exactly the same method as the NSVZ $\beta$-function in the case of rigid ${\cal N}=1$ SQED \cite{Nartsev:2016nym}. For RGF defined in terms of the renormalized coupling constant this relation is also valid in the $\mbox{HD}+\mbox{MSL}$ scheme \cite{Nartsev:2016mvn}.

\section{Adler $D$-function in ${\cal N}=1$ SQCD}
\hspace{\parindent}

NSVZ-like expression can be also written for the Adler $D$-function \cite{Adler:1974gd} in (massless) ${\cal N}=1$ SQCD interacting with the Abelian gauge field \cite{Shifman:2014cya,Shifman:2015doa},

\begin{eqnarray}
&&\hspace*{-1cm} S =
\frac{1}{2 g_{0}^2}\mbox{tr}\,\mbox{Re} \int d^4x\, d^2\theta\,W^a W_a + \frac{1}{4 e_0^2} \mbox{Re}
\int d^4x\, d^2\theta\, \bm{W}^a \bm{W}_a\nonumber\\
&&\hspace*{-1cm} + \sum\limits_{f=1}^{N_f}\, \frac{1}{4} \int d^4x\, d^4\theta\Big(\phi_f^+ e^{2 q_f \bm{V}+2V}\phi_f + \widetilde\phi_f^+ e^{-2 q_f \bm{V} -2V^t}
\widetilde\phi_f\Big).
\end{eqnarray}

\noindent
This theory is invariant under the $SU(N_c)\times U(1)$ gauge transformations. The chiral matter superfields $\phi_f$ and $\widetilde\phi_f$ belong to the fundamental representation of $SU(N_c)$ and have the charges $q_f e$ and $-q_f e$ with respect to the group $U(1)$, respectively. In our notation $V$ is the non-Abelian $SU(N_c)$ gauge superfield and ${\bm V}$ is the Abelian $U(1)$ gauge superfield. Evidently, the theory contains two coupling constants, $\alpha_s = g^2/4\pi$ and $\alpha = e^2/4\pi$.

The $D$-function encodes quantum corrections to the electromagnetic coupling constant $\alpha$ which appear due to the strong interaction. In the supersymmetric case this implies that the electromagnetic gauge superfield $\bm{V}$ is treated as an external field. Due to the Ward identity the two-point Green function of this superfield is transversal,

\begin{equation}
\Delta\Gamma^{(2)} = - \frac{1}{16\pi} \int \frac{d^4p}{(2\pi)^4}\,d^4\theta\,\mbox{\boldmath$V$}\,\partial^2\Pi_{1/2} \mbox{\boldmath$V$}\,
\Big(d^{-1}(\alpha_0,\alpha_{0s},\Lambda/p)-\alpha_0^{-1}\Big).
\end{equation}

\noindent
The Adler function can be defined in terms of the bare coupling constant by the equation

\begin{equation}
D(\alpha_{0s}) = \frac{3\pi}{2} \frac{d}{d\ln\Lambda} \Big(d^{-1}(\alpha_0,\alpha_{0s},\Lambda/p) - \alpha_0^{-1}\Big)\Big|_{p=0} = \frac{3\pi}{2\alpha_0^2}
\frac{d\alpha_0}{d\ln\Lambda}.
\end{equation}

\noindent
Again, this function depends on regularization, but is independent of a renormalization prescription for a fixed regularization.

According to \cite{Shifman:2014cya,Shifman:2015doa}, in the case of using the higher covariant derivative regularization\footnote{The higher derivative term for the considered theory should contain covariant derivatives $\nabla_a = e^{-\Omega^+} D_a e^{\Omega^+}$; $\bar\nabla_{\dot a} = e^{\Omega} \bar D_{\dot a} e^{-\Omega}$, where $e^{2V} = e^{\Omega^+} e^\Omega$.} the exact expression for the Adler function for the considered theory can be written in the NSVZ-like form

\begin{equation}\label{Exact_Adler}
D(\alpha_{0s}) = \frac{3}{2} \sum\limits_f q_f^2\cdot N_c \Big(1-\gamma(\alpha_{0s})\Big).
\end{equation}

\noindent
It looks very similar to the NSVZ $\beta$-function in ${\cal N}=1$ SQED and is derived in all loops by exactly the same method. However, Eq. (\ref{Exact_Adler}) contains the anomalous dimension of the {\it non-Abelian} theory, and this is a very essential difference from the ${\cal N}=1$ SQED case. Recently this expression has been confirmed by an explicit three-loop calculation in Ref. \cite{Kataev:2017qvk}.

\section{Non-Abelian ${\cal N}=1$ supersymmetric theories}

\subsection{Regularization and renormalization}
\hspace{\parindent}

Let us consider the theory described by the action (\ref{SYM_Superfield_Action}) in the massless limit. It is convenient to do calculations using the background field method introduced by replacement $e^{2V} \to e^{\bm{\Omega}^+} e^{2V} e^{\bm{\Omega}}$. The background gauge superfield $\bm{V}$ is then related to $\bm{\Omega}$ and $\bm{\Omega}^+$ by the equation $e^{2\bm{V}} = e^{\bm{\Omega}^+} e^{\bm{\Omega}}$. The higher derivative term in this case can be written in the form

\begin{eqnarray}
&&\hspace*{-5mm} S_{\Lambda} =
\frac{1}{2e_0^2}\,\mbox{Re}\,\mbox{tr}\int d^4x\, d^2\theta\,
e^{\Omega} e^{\bm{\Omega}} W^a e^{-\bm{\Omega}} e^{-\Omega}
\Big[R\Big(-\frac{\bar\nabla^2
\nabla^2}{16\Lambda^2}\Big) -1\Big]_{Adj} e^{\Omega} e^{\bm{\Omega}} \qquad\nonumber\\
&&\hspace*{-5mm} \times W_a e^{-\bm{\Omega}} e^{-\Omega} + \frac{1}{4} \int d^4x\,
d^4\theta\,\phi^+ e^{\bm{\Omega}^+} e^{\Omega^+}
\Big[F\Big(-\frac{\bar\nabla^2 \nabla^2}{16\Lambda^2}\Big)-1\Big]
e^{\Omega} e^{\bm{\Omega}} \phi,
\end{eqnarray}

\noindent
where the functions $R(x)$ and $F(x)$ rapidly increase at infinity and satisfy the condition $R(0)=F(0)=1$. It is convenient to fix a gauge without breaking the background gauge invariance. For this purpose it is possible to use the gauge fixing term

\begin{equation}
S_{\mbox{\scriptsize gf}} = -\frac{1}{16\xi_0 e_0^2} \mbox{tr}\int
d^4x\, d^4\theta\, \bm{\nabla}^2 V K\Big(-\frac{\bm{\bar\nabla}^2
\bm{\nabla}^2}{16\Lambda^2}\Big)_{Adj} \bm{\bar\nabla}^2 V,
\end{equation}

\noindent
where $K(0)=1$ and $K(x)$ also rapidly grows at infinity. The corresponding actions for ghosts and the Pauli--Villars determinants can be found in Ref. \cite{Aleshin:2016yvj}, where they are discussed in all details. The renormalization constants are introduced by the equations

\begin{equation}\label{Z_Definition}
\frac{1}{\alpha_0} =
\frac{Z_\alpha}{\alpha};\quad V = Z_V Z_\alpha^{-1/2} V_R;\quad \bar c c =
Z_c Z_\alpha^{-1} \bar c_R c_R;\quad \phi_i =
(\sqrt{Z_\phi})_i{}^j (\phi_R)_j,
\end{equation}

\noindent
where $\bar c$ and $c$ are the chiral Faddeev--Popov ghost superfields.

\subsection{Finiteness of the triple gauge-ghost vertices}
\hspace{\parindent}

In ${\cal N}=1$ gauge supersymmetric theories the three-point gauge-ghost vertices ($\bar c\, V c$, $\bar c^+ V c$, $\bar c\, V c^+$, and $\bar c^+ V c^+$) with two ghost legs and a single leg of the quantum gauge superfield are finite \cite{Stepanyantz:2016gtk}, so that

\begin{equation}\label{Z_VCC}
\frac{d}{d\ln\Lambda} (Z_\alpha^{-1/2} Z_c Z_V) = 0.
\end{equation}

\noindent
(At the one-loop level it was found in \cite{Aleshin:2016yvj}.) This theorem is derived by the help of the Slavnov--Taylor identities, which can be obtained using the standard methods \cite{Taylor:1971ff,Slavnov:1972fg}. To write the identity for the considered three-point functions, we introduce the chiral source ${\cal J}$ and the source term

\begin{equation}
-\frac{e_0}{2} \int d^4x\, d^2\theta\, f^{ABC} {\cal J}^A c^B c^C +\mbox{c.c.}
\end{equation}

\noindent
Then using the superspace Feynman rules it is possible to prove that the effective vertex

\begin{equation}\label{H_Definition}
\frac{\delta^3\Gamma}{\delta c_z^C \delta c_w^D \delta {\cal J}_y^B}
= \frac{e_0}{4} f^{BCD} \int \frac{d^4p}{(2\pi)^4} \frac{d^4q}{(2\pi)^4} H(p,q)
\bar D_{z}^2\delta^8_{zy}(q+p) \bar D_{w}^2 \delta^8_{yw}(q)
\end{equation}

\noindent
is finite in all orders. Really, we can present the corresponding superdiagrams as integrals over the total superspace, which include integration over

\begin{equation}
\int d^4\theta = -\frac{1}{2} \int d^2\theta \bar D^2 +\ \mbox{total derivatives in the coordinate space}.
\end{equation}

\noindent
Consequently, due to chirality of all external legs the non-vanishing result can be obtained only if two right spinor derivatives also act to the external legs. Thus, commuting supersymmetric covariant derivatives, we see that the result should be proportional to, at least, second degree of the external momenta and is finite in the ultraviolet region.

From dimensional and chirality considerations one can write the following expression for the one of triple gauge-ghost Green functions,

\begin{eqnarray}\label{Three-Point_Function1}
&&\hspace*{-5mm} \frac{\delta^3\Gamma}{\delta \bar c_x^{*A} \delta V_y^B \delta c_z^C} = -\frac{i e_0}{16} f^{ABC} \int \frac{d^4p}{(2\pi)^4} \frac{d^4q}{(2\pi)^4} \Big(f(p,q) \partial^2\Pi_{1/2} \nonumber\\
&&\hspace*{-5mm} - F_\mu(p,q) (\gamma^\mu)_{\dot a}{}^{b} \bar D^{\dot a} D_b + F(p,q) \Big)_{y} \Big(D_{x}^2\delta^8_{xy}(q+p)\, \bar D_{z}^2 \delta^8_{yz}(q)\Big),
\end{eqnarray}

\noindent
where $\delta^8_{xy}(p) \equiv \delta^4(\theta_x-\theta_y) e^{ip_\alpha (x^\alpha - y^\alpha)}$. Then the Slavnov--Taylor identity can be written in the form

\begin{equation}\label{STI1}
G_c(q) F(q,p) + G_c(p) F(p,q) = 2 G_c(q+p) H(-q-p,q),
\end{equation}

\noindent
where $G_c(q)$ is the two-point Green function for the Faddeev--Popov ghosts. Multiplying this equation to $Z_c$, differentiating the result with respect to $\ln\Lambda$ and setting $p=-q$ we obtain finiteness of the function $F(-q,q)$, which follows from the finiteness of $(G_c)_{R}$ and $H$ in the limit $\Lambda\to \infty$. This means that the corresponding renormalization constant is finite, see Eq. (\ref{Z_VCC}). Consequently, all three-point ghost-gauge vertices are also finite.

\subsection{$V\bar c c$-vertices in the one-loop approximation}
\hspace{\parindent}

In the one-loop approximation (after the Wick rotation)

\begin{eqnarray}\label{F_Explcit}
&&\hspace*{-6mm} F(p,q) = 1 + \frac{e_0^2 C_2}{4} \int \frac{d^4k}{(2\pi)^4} \Bigg\{-\frac{(q+p)^2}{R_k k^2 (k+p)^2 (k-q)^2} - \frac{\xi_0\, p^2}{K_k k^2 (k+q)^2}\nonumber\\
&&\hspace*{-6mm} \times \frac{1}{(k+q+p)^2}
+ \frac{\xi_0\, q^2}{K_k k^2 (k+p)^2 (k+q+p)^2} + \Big(\frac{\xi_0}{K_k}
- \frac{1}{R_k}\Big)\Bigg(-\frac{1}{k^2 (k+q)^2} \nonumber\\
&&\hspace*{-6mm} - \frac{1}{k^2 (k+p)^2} + \frac{2}{k^2 (k+q+p)^2} - \frac{2(q+p)^2}{k^4 (k+q+p)^2}\Bigg)\Bigg\} + O(\alpha_0^2,\alpha_0\lambda_0^2).
\end{eqnarray}

\noindent
It is easy to see that this expression is finite in the UV region. The other functions in Eq. (\ref{Three-Point_Function1}) are also finite, see \cite{Stepanyantz:2016gtk}. The finiteness of the function $H$, defined in Eq. (\ref{H_Definition}), at the one-loop level has also been demonstrated,

\begin{eqnarray}
&&\hspace*{-6mm} H(p,q) = 1 - \frac{e_0^2 C_2}{4} \int \frac{d^4k}{(2\pi)^4}
\Bigg\{\frac{p^2}{R_k k^2 (k+q)^2 (k+q+p)^2}  + \frac{(q+p)^2}{k^4 (k+q+p)^2}\nonumber\\
&&\hspace*{-6mm} \times \Big(\frac{\xi_0}{K_k} - \frac{1}{R_k}\Big) + \frac{q^2}{k^4 (k+q)^2} \Big(\frac{\xi_0}{K_k} - \frac{1}{R_k}\Big) \Bigg\} + O(e_0^4, e_0^2 \lambda_0^2).\vphantom{\Bigg)}
\end{eqnarray}

\subsection{New form of the NSVZ relation}
\hspace{\parindent}

Let write the NSVZ relation (\ref{NSVZ_Beta_Function}) for RGF defined in terms of the bare couplings (see the definitions in Ref. \cite{Stepanyantz:2016gtk}) in the form

\begin{equation}\label{NSVZ_Equivalent}
\frac{\beta(\alpha_0,\lambda_0)}{\alpha_0^2} = - \frac{3 C_2 - T(R)
+ C(R)_i{}^j (\gamma_\phi)_j{}^i(\alpha_0,\lambda_0)/r}{2\pi}
+ \frac{C_2}{2\pi}\cdot \frac{\beta(\alpha_0,\lambda_0)}{\alpha_0}
\end{equation}

\noindent
and take into account that the $\beta$-function can be related to the renormalization constant $Z_\alpha$,

\begin{equation}
\beta(\alpha_0,\lambda_0) = \frac{d\alpha_0(\alpha,\lambda,\Lambda/\mu)}{d\ln\Lambda}\Big|_{\alpha,\lambda=\mbox{\scriptsize const}}
= -\alpha_0 \frac{d\ln Z_\alpha}{d\ln\Lambda}\Big|_{\alpha,\lambda=\mbox{\scriptsize const}}.
\end{equation}

\noindent
Then the right hand side of Eq. (\ref{NSVZ_Equivalent}) can be expressed in terms of $\gamma_c$ and $\gamma_V$ by the help of Eq. (\ref{Z_VCC}),

\begin{equation}
\beta(\alpha_0,\lambda_0)
= -2\alpha_0 \frac{d\ln (Z_c Z_V)}{d\ln\Lambda}\Big|_{\alpha,\lambda=\mbox{\scriptsize const}}
= 2\alpha_0 \Big(\gamma_c(\alpha_0,\lambda_0) + \gamma_V(\alpha_0,\lambda_0)\Big).
\end{equation}

\noindent
Substituting this identity into Eq. (\ref{NSVZ_Equivalent}) we rewrite the exact NSVZ $\beta$-function in a different form,

\begin{eqnarray}\label{NSVZ_New_Form}
&& \frac{\beta(\alpha_0,\lambda_0)}{\alpha_0^2} = - \frac{1}{2\pi}\Big(3 C_2 - T(R) - 2C_2 \gamma_c(\alpha_0,\lambda_0) - 2C_2 \gamma_V(\alpha_0,\lambda_0)\qquad\nonumber\\
&& + C(R)_i{}^j (\gamma_\phi)_j{}^i(\alpha_0,\lambda_0)/r\Big).
\end{eqnarray}

\noindent
Eq. (\ref{NSVZ_New_Form}) admits a simple graphical interpretation similar to the Abelian case. Consider a supergraph without external lines. By attaching two external legs of the superfield $\bm{V}$ we obtain a set of diagrams contributing to the $\beta$-function. From the other side, cutting internal lines gives superdiagrams contributing to the anomalous dimensions of the Faddeev--Popov ghosts, of the quantum gauge superfield, and of the matter superfields. Eq. (\ref{NSVZ_New_Form}) relates these two sets of superdiagrams.

\subsection{The NSVZ scheme for non-Abelian gauge theories}
\hspace{\parindent}

The RGF standardly defined in terms of the renormalized couplings (we again denote them by tildes) are scheme-dependent and satisfy the NSVZ relation only in a certain (NSVZ) subtraction scheme. Let us suggest that, similar to the Abelian case, RGF defined in terms of the bare couplings satisfy the NSVZ relation (\ref{NSVZ_New_Form}) in the case of using the higher covariant derivative regularization. Really, the qualitative way of its derivation looks exactly as in ${\cal N}=1$ SQED and the factorization into total derivatives \cite{Pimenov:2009hv,Steklov} and double total derivatives \cite{Stepanyantz:2011bz} also takes place at least in the lowest orders. Then, repeating the argumentation of Ref. \cite{Kataev:2013eta}, one can prove that in the non-Abelian case both definitions of RGF give the same result (for coinciding arguments) if the renormalization constants satisfy the conditions

\begin{equation}\label{NSVZ_Conditions}
Z_\alpha(\alpha,\lambda,x_0) = 1;\qquad
(Z_\phi)_i{}^j(\alpha,\lambda,x_0)=\delta_i{}^j;\qquad
Z_c(\alpha,\lambda,x_0)=1.
\end{equation}

\noindent
Thus, under the assumption that the NSVZ relation is valid for RGF defined in terms of the bare couplings with the higher derivative regularization, the NSVZ scheme is given by the boundary conditions (\ref{NSVZ_Conditions}). Again, it is easy to see that for $x_0=0$ in this scheme only powers of $\ln\Lambda/\mu$ are included into the renormalization constants, so that the NSVZ scheme coincides with $\mbox{HD}+\mbox{MSL}$. Certainly, it is also assumed that $Z_V = Z_\alpha^{1/2} Z_c^{-1}$ due to the non-renormalization of the $V\bar c c$-vertices.

\subsection{Checking the new form of the NSVZ relation by explicit calculations}
\hspace{\parindent}

\begin{figure}[h]
\begin{picture}(0,2)
\put(3.8,0){\includegraphics[scale=0.07]{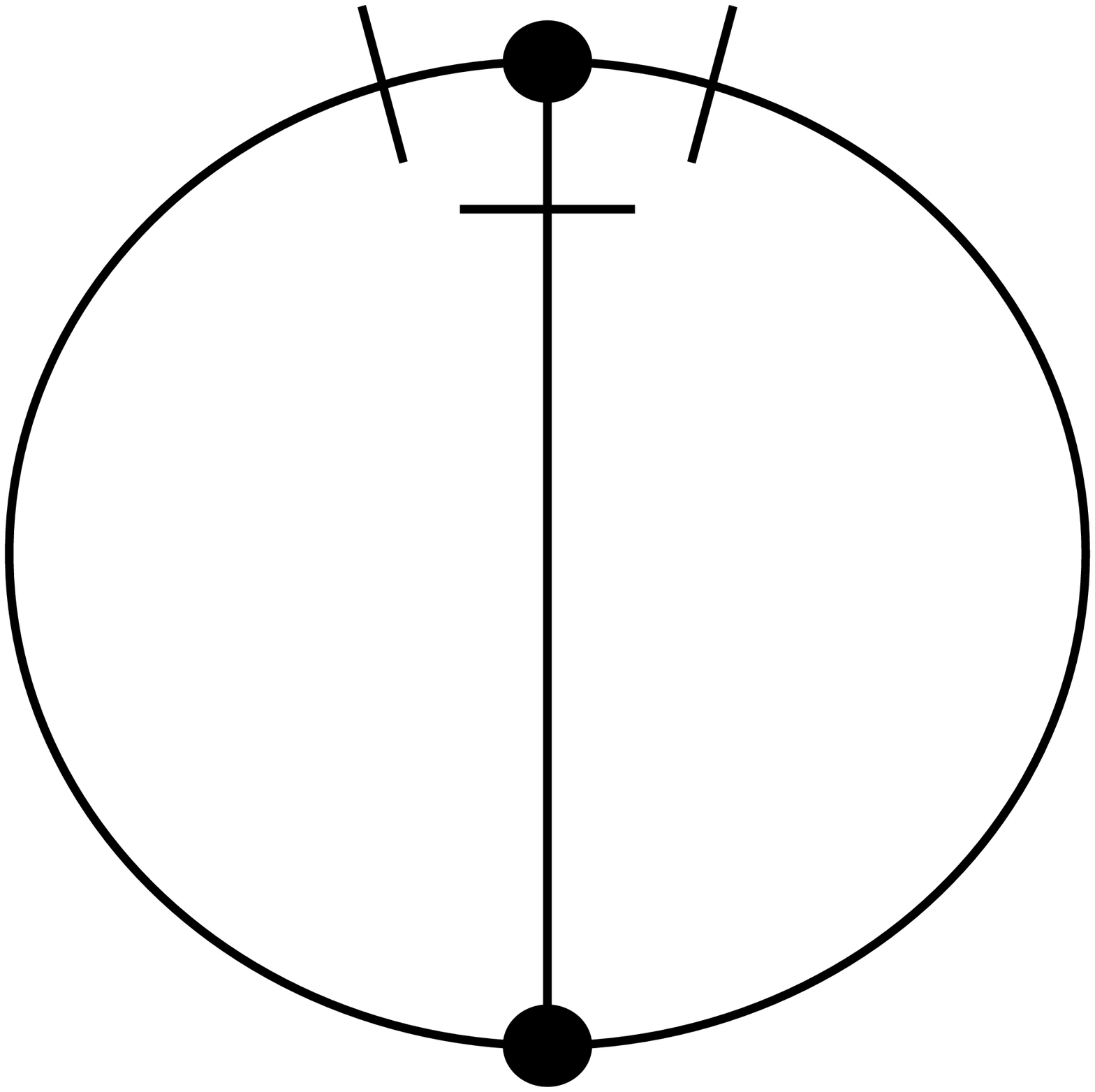}}
\put(7.8,0){\includegraphics[scale=0.07]{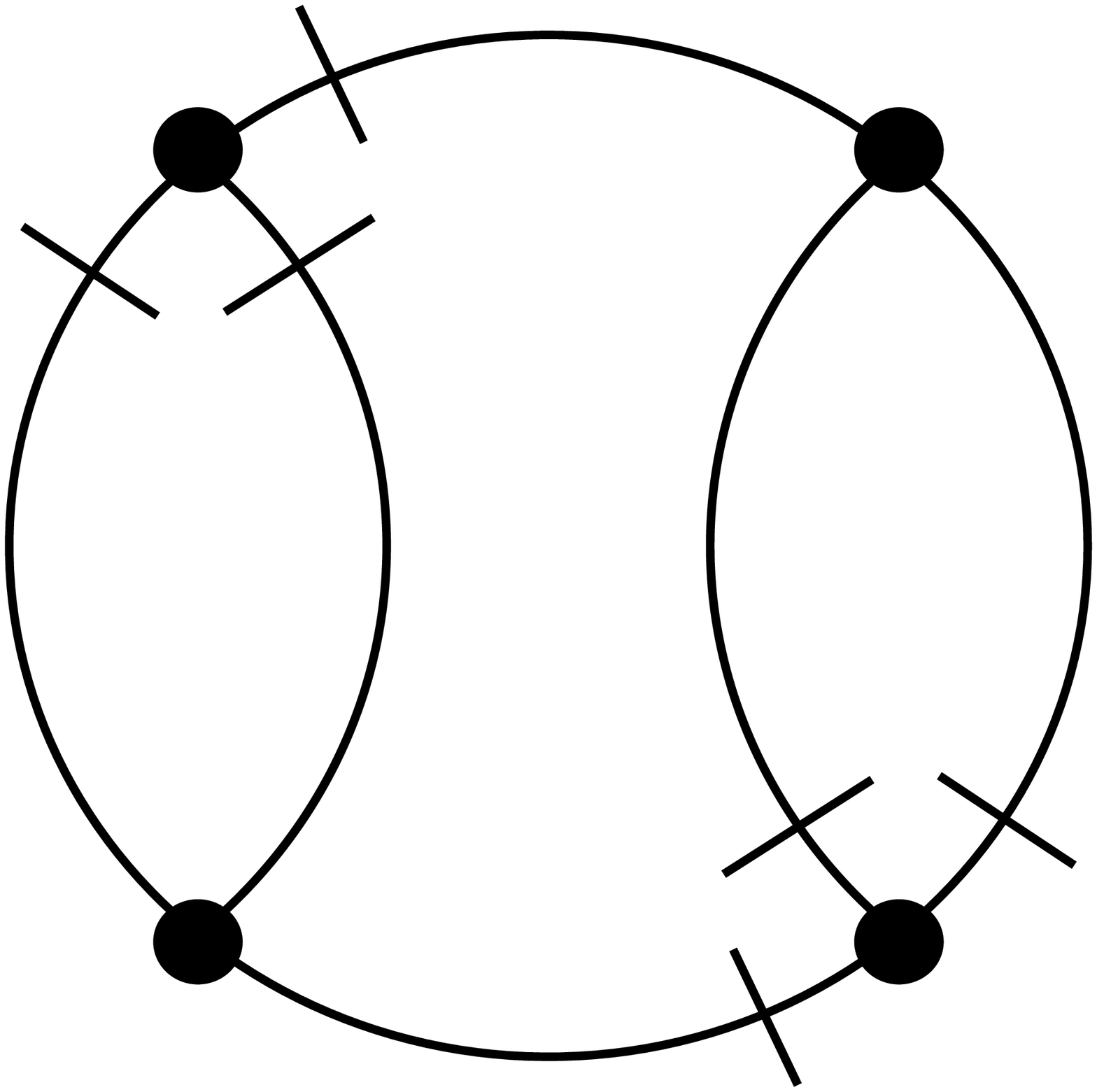}}
\end{picture}
\caption{The terms in the NSVZ relation which are investigated here are obtained from these two graphs.}\label{Figure_Graphs}
\end{figure}

To check the above results, we consider terms quartic in the Yukawa couplings \cite{Shakhmanov:2017soc} corresponding to the graphs presented in Fig. \ref{Figure_Graphs}.

Attaching two external lines of the background gauge superfield gives a large number of two- and three-loop diagrams contributing to the $\beta$-function. The corresponding diagrams for the anomalous dimension are obtained by cutting internal lines in the considered graphs. The result for the considered part of the $\beta$-function defined in terms of the bare couplings can be presented as an integral of double total derivatives,

\begin{eqnarray}\label{Delta_Beta}
&&\hspace*{-5mm} \frac{\Delta\beta(\alpha_0,\lambda_0)}{\alpha_0^2} = -\frac{2\pi}{r} C(R)_i{}^j \frac{d}{d\ln\Lambda} \int \frac{d^4k}{(2\pi)^4} \frac{d^4q}{(2\pi)^4} \,\lambda_0^{imn} \lambda^*_{0jmn} \frac{\partial}{\partial q_\mu} \frac{\partial}{\partial q^\mu}\nonumber\\
&&\hspace*{-5mm} \times \Big(\frac{1}{k^2 F_k\, q^2 F_q\, (q+k)^2 F_{q+k}}\Big)
+ \frac{4\pi}{r} C(R)_i{}^j \frac{d}{d\ln\Lambda} \int \frac{d^4k}{(2\pi)^4} \frac{d^4l}{(2\pi)^4} \frac{d^4q}{(2\pi)^4}\nonumber\\
&&\hspace*{-5mm} \times \Bigg(\lambda_0^{iab}\lambda^*_{0kab} \lambda_0^{kcd}\lambda^*_{0jcd} \Big(\frac{\partial}{\partial k_\mu} \frac{\partial}{\partial k^\mu} - \frac{\partial}{\partial q_\mu} \frac{\partial}{\partial q^\mu}\Big) + 2 \lambda_0^{iab}\lambda^*_{0jac}
\lambda_0^{cde}\lambda^*_{0bde}\,\nonumber\\
&&\hspace*{-5mm} \times \frac{\partial}{\partial q_\mu} \frac{\partial}{\partial q^\mu} \Bigg) \frac{1}{k^2 F_k^2\, q^2 F_q\, (q+k)^2 F_{q+k}\, l^2 F_l\, (l+k)^2 F_{l+k}}.
\end{eqnarray}

\noindent
Taking one of loop integrals it is possible to relate this expression to the corresponding contribution to the anomalous dimension of the matter superfield (defined in terms of the bare couplings),

\begin{equation}
\frac{\Delta\beta(\alpha_0,\lambda_0)}{\alpha_0^2} = -\frac{1}{2\pi r} C(R)_i{}^j \Delta\gamma_\phi(\lambda_0)_j{}^i.
\end{equation}

\noindent
This equation completely agrees with Eq. (\ref{NSVZ_New_Form}), so that the NSVZ relation is satisfied for terms of the considered structure.

For $F(k^2/\Lambda^2) = 1+ k^2/\Lambda^2$ all loop integrals can be calculated,

\begin{equation}
\Delta\gamma_\phi(\alpha_0,\lambda_0)_j{}^i = \frac{1}{4\pi^2} \lambda_0^{iab} \lambda^*_{0jab}
- \frac{1}{16\pi^4}{\lambda_0^{iab} \lambda^*_{0jac} \lambda_0^{cde} \lambda^*_{0bde}}.
\end{equation}

\noindent
Scheme-dependent RGF defined in terms of the renormalized couplings have been calculated in Ref. \cite{Shakhmanov:2017soc}. The contribution to the $\beta$-function depends on some finite constants $g_1$ and $b_2$, which appear due to arbitrariness of choosing a subtraction scheme,

\begin{eqnarray}
&&\hspace*{-9mm} \widetilde\gamma_\phi(\alpha,\lambda)_j{}^i = \frac{1}{4\pi^2} \lambda^{iab} \lambda^*_{jab}
- \frac{1}{16\pi^4} \lambda^{iab} \lambda^*_{jac} \lambda^{cde} \lambda^*_{bde} + O(\alpha) + O(\lambda^6);\\
&&\hspace*{-9mm} \frac{\widetilde\beta(\alpha,\lambda)}{\alpha^2} =  -\frac{1}{2\pi}\Big(3C_2 - T(R)\Big) + \frac{1}{2\pi r} C(R)_i{}^j \Big[ -\frac{1}{4\pi^2} \lambda^{iab} \lambda^*_{jab} + \frac{1}{16\pi^4}\nonumber\\
&&\hspace*{-9mm} \times \lambda^{iab} \lambda^*_{kab}  \lambda^{kcd} \lambda^*_{jcd} \Big(b_2 -g_1\Big)
+ \frac{1}{16\pi^4} \lambda^{iab} \lambda^*_{jac} \lambda^{cde} \lambda^*_{bde} \Big(1+ 2b_2 -2g_1\Big)\Big]\nonumber\\
&&\hspace*{-9mm} + O(\alpha) + O(\lambda^6).\qquad\vphantom{\frac{1}{2}}
\end{eqnarray}

\noindent
We see that for an arbitrary values of $g_1$ and $b_2$ the NSVZ relation is not valid. However, the values of $g_1$ and $b_2$ can be fixed by imposing the conditions (\ref{NSVZ_Conditions}). In this case $g_1 = b_2 = -x_0$, so that $b_2 - g_1 = 0$. Therefore, in this scheme

\begin{equation}
\frac{\widetilde\beta(\alpha,\lambda)}{\alpha^2}
= -\frac{1}{2\pi}\Big(3C_2 - T(R)\Big) - \frac{1}{2\pi r} C(R)_i{}^j \widetilde\gamma_\phi(\alpha,\lambda)_i{}^j + O(\alpha) + O(\lambda^6).
\end{equation}

\noindent
This confirms the guess that Eq. (\ref{NSVZ_Conditions}) gives the NSVZ scheme in the non-Abelian case.

Note that recently \cite{Shakhmanov:2017wji} the identity (\ref{NSVZ_New_Form}) has been completely checked in the two-loop approximation in the case of using the non-invariant version of the higher covariant derivative regularization supplemented by a special subtraction procedure which restores the Slavnov--Taylor identities \cite{Slavnov:2003cx}.

\section{Conclusion}
\hspace{\parindent}

The $\beta$-function defined in terms of the bare coupling constant for ${\cal N}=1$ supersymmetric gauge theories regularized by higher derivatives is given by integrals of double total derivatives. In some cases it has been proved in all loops, but for general non-Abelian SYM theories at present there are only strong evidences in favour of this. Such a structure of quantum corrections naturally leads to the NSVZ relation for RGF defined in terms of the bare coupling constant, which is obtained after taking the integral of the total derivative and is valid independently of the subtraction scheme. Note that in the non-Abelian case an important ingredient of the derivation is the finiteness of the three-point ghost-gauge vertices, which allows rewriting the NSVZ equation in a different form.

The RGF defined in terms of the renormalized couplings satisfy the NSVZ relation only in a certain (NSVZ) scheme, which is obtained with the higher derivative regularization by minimal subtraction of logarithms. This means that only powers of $\ln\Lambda/\mu$ are included into various renormalization constants. This prescription can be also reformulated by imposing simple boundary conditions on the renormalization constants.

All general statements considered here are confirmed by explicit perturbative calculations. Note that some of them are made in the three-loop approximation and are highly non-trivial.

\section*{Acknowledgements}

I am very grateful to the organizers of the twentieth workshop "What Comes Beyond the Standard Model" (July 9-17, 2017, Bled, Slovenia) for kind hospitality and to all participants for very helpful discussions. Also I would like to express my deep thanks to A.L.Kataev for numerous discussions of various topics considered in this paper.

\end{document}